\documentclass[10pt]{article}

\usepackage[a4paper,textwidth=18cm,textheight=24cm,top=2.85cm, bottom=2.85cm, left=1.5cm, right=1.5cm]{geometry}
\pdfoutput=1
\usepackage{scicite}
\usepackage{hyperref}
\usepackage{icdp2009}
\usepackage{lmodern}
\usepackage{graphicx, amsmath}
\usepackage{times}

\makeatletter
\long\def\@makecaption#1#2{%
\vskip\abovecaptionskip
\sbox\@tempboxa{#1. #2}%
\ifdim \wd\@tempboxa >\hsize
#1. #2\par
\else
\global \@minipagefalse
\hb@xt@\hsize{\box\@tempboxa\hfil}%
\fi
\vskip\belowcaptionskip}
\makeatother

\newcommand{\fref}[1]{Figure~(\ref{#1})}
\newcommand{\eref}[1]{Equation~\eqref{#1}}
\newcommand{\xpar}{\\\par\noindent}
\newcommand{\ypar}{\par\noindent}
\begin{document}
\noindent

\bibliographystyle{ieeetr}

\title{An Automated Global Method for Extraction of Distance Distributions from Electron Spin Resonance Pulsed Dipolar Signals}

\authorname{Aritro Sinha Roy$^{1, \dagger}$}
\authoraddr{\small $^1$Department of Chemistry \& Chemical Biology \\ \small Cornell University, Ithaca, New York, USA \\ \small $^\dagger$\emph{Work in progress: preparing manuscript from PhD thesis work, adviser's name to be added in the final draft}}


\maketitle

\abstract
Electron spin resonance (ESR) pulsed dipolar spectroscopy (PDS) is used effectively in measuring nano-meter range distances for protein structure prediction. The current global approach in extracting the distance distribution from time domain PDS signal has multiple limitations. We present a parameter free global method, which is more efficacious and less sensitive to signal noise compared to the current method.


\section*{Introduction}
ESR pulsed dipolar spectroscopy can measure intramolecular distances in the range of 1-10 nm with high accuracy in bi-labeled protein molecules, and plays a key role in protein 3D structure prediction \cite{Freed_Science,Jeschke_Book,deer_rev,epr_methods}. However, extraction of the distance distribution, P(r) from the PDS time domain signal is an \emph{ill-posed} problem \cite{ill_posed_1,ill_posed_2}. Tikhonov regularization is used currently as the global method to solve the problem \cite{tkn_acert,tkn_1,tkn_2}, but it is highly sensitive to the following signal attributes
\begin{itemize}
\item[1.] signal-to-noise ratio (snr)
\item[2.] dipolar evolution time, $t_m$
\item[3.] nature of the distance distribution
\end{itemize}
It is challenging and costly to acquire signals for sufficiently long $t_m$, and with high snr, while the nature of the distribution is often complex. Other global approaches such as model fitting \cite{deer_gaussian1,deer_gaussian2} and DEERNet \cite{deer_dnn} are developed. These methods require priori model, and large training, respectively and hence, are not generalizable. The Srivastava-Freed Singular Value Decomposition (SF-SVD) method \cite{svd_1, svd_2} is a recently developed local approach, which effectively extracts distance distributions from PDS signals. It is necessary to develop a global method, which is less sensitive to the snr, the dipolar evolution time, and the nature of the distribution to complement such efficient local approaches in deriving the unique distance distribution.
\section*{Theoretical Background}
The integral form of the PDS signal is given by
\begin{equation}
S(t) = \int_{r_{min}}^{r_{max}}{\kappa{(t, r)}\,P(r)\,dr}
\label{eqn:fredholm}
\end{equation}
where S(t) is the time domain signal, $\kappa$ is the pairwise dipolar interaction kernel at a given $r$ averaged over all possible orientations, and P(r) is the distance distribution. \eref{eqn:fredholm} is an example of Fredholm's integral of the first kind \cite{fredholm}. For the purpose of signal analysis, the discrete form of the signal can be written as
\begin{equation}
S = \omega\,\kappa\,P
\label{eqn:fredholm_discrete}
\end{equation}
Numerical coefficient, $\omega$ stems from discretization of the signal \cite{num_anal}. P cannot be obtained by an inversion of $\kappa$, because $\kappa$ is almost always singular and hence, solving for P from \eref{eqn:fredholm_discrete} is an \emph{ill-posed} problem \cite{ill_posed_1,ill_posed_2}. The current modus operandi in deriving P(r) from PDS signal is Tikhonov regularization \cite{tkn_acert,tkn_1,tkn_2}, represented by the penalized minimization
\begin{equation}
P = min\big(\|S-\kappa\,P_\lambda\|^2\, + \, \lambda^2\,\|L\,P_\lambda\|\big)
\label{eqn:tikhonov_1}
\end{equation}
The distribution obtained from \eref{eqn:tikhonov_1} is given by
\begin{equation}
P_\lambda = (\kappa^T\,\kappa + \lambda^2\,L^T\,L)^{-1}\,\kappa^T\,S
\label{eqn:pr_tkn}
\end{equation}
where $\lambda$ is the regularization parameter and $L$ is the regularization operator. Three reasonable choices of $L$ are the identity matrix ($L_0$), the first derivative ($L_1$) and the second derivative ($L_2$). In this work, we have used $L_0$ and the value of $\lambda$ is obtained using the L-curve maximum curvature method \cite{tkn_acert}. The L-curve is a log-log plot of the monotonically decaying Tikhonov penalty term, $\eta = \|L\,P_\lambda\|$ against the residual norm, $\rho = \|S-\kappa\,P_\lambda\|$. The $\lambda$ value corresponding to the corner of the L-curve or the position of the maximum positive curvature is taken as its optimal value and it is given by \cite{opt_tkn}
\begin{equation}
\lambda_{mc} = argmax\,\frac{\hat{\rho}'\,\hat{\eta}'' - \hat{\eta}'\,\hat{\rho}''}{(\hat{\rho}'^2 + \hat{\eta}'^2)^{3/2}}
\label{eqn:max_curve}
\end{equation}
where $\hat{\eta} = \log{(\eta)}$ and $\hat{\rho} = \log{(\rho)}$. Even though there are various other methods available to find the value of $\lambda$, the challenge remains in finding the \emph{true optimal value} that causes neither under, or overfitting \cite{opt_tkn}.
\section*{ExPDS: A Global Method of P(r) Derivation}
In this work, we present a novel method, ExPDS (\textbf{Ex}tranction of information from \textbf{PDS} signals) for deriving P(r) from PDS signals in a fully automated fashion and it is written in Python script to facilitate distribution, and user access. ExPDS is a parameter free method, which employs a modified stochastic gradient descent (SGD) algorithm \cite{sgd_rev} for the minimization problem presented in \eref{eqn:tikhonov_1}. In the ordinary gradient descent method, a function is minimized by iteratively proceeding in the direction of the steepest descent. In order to optimize a function, $h = X\cdot{P}$, which is the reconstructed signal in our case, the $N\times{n}$ design matrix, $X$ is constructed by evaluating the dipolar kernel at each time increment ($N$ values), and $r$ ($n\le{N}$ values). The initial P(r) is taken to be a null vector and it is updated at the end of each iteration constrained to P$\ge0$ as follows
\begin{align*}
&h = X\cdot{P} \\
&cost function = \|S - h\| \\
&k^{th} gradient, \Delta_k = \frac{\alpha}{N}\,\left(-X_k\,(S - h)\right)\,\,\,\,\,\,k\le{n} \\
&\textrm{($\alpha$ is the learning rate)} \\
&P_k = P_k - \Delta_k \\
&\textrm{constrained to $P_k\ge{0}$} \\
\end{align*}
The learning rate, $\alpha$ is the rate at which the calculation moves along the steepest descent. In each iteration, (1) $\alpha$ is selected randomly from a range of values, and (2) a random, equally spaced "mini-batch" of the PDS time domain signal is selected in running 10 iterations of the gradient descent minimization. Strategy (1) eliminates the need of finding an optimum value of $\alpha$, and ensures that the calculation does not get trapped in a local minima \cite{avg_sgd}. We note that any segment of the PDS signal can reproduce the distance distribution entirely or partially. If the noise is truly random, its effect on the derived distribution from the various segments of the signal is expected to be random and hence, in updating P(r) by strategy (2) suppresses such random contributions significantly.
\xpar The default range of $\alpha$ is set to $\{0.0001, 0.1\}$ and the mean squared error (mse) is used as the metric of convergence. If the mse does not decay monotonically, we infer that the rate of descendence is too high, or in other words, the range of $\alpha$ should be lowered. Hence, in such cases, the maximum value of $\alpha$ is reduced by a factor of 2 and the process is continued till the convergence criteria is satisfied, or a monotonically decaying mse is obtained. This entire operation is free of any parameter tuning or human intervention.
\xpar We denote the optimal range of $\alpha$ as $R_\alpha$, and the corresponding solution, and mse as $P^{A}$, and $mse^{A}$, respectively. A lowered range, $R_\alpha/10$ let the calculation sweep a larger solution space and yields an over-fitted solution, which is used to mark the confidence interval (CI). Following that, the sensitivity analysis is conducted by varying each value of the solution $P_{i}^{A}$ between $0.25\,P_{i}^{A}$ and $4.0\,P_{i}^{A}$, keeping all the other values in $P^A$ fixed, and the value corresponding to the minimum mean squared error is recorded as an alternative set of solution, $P_{i}^{B}$. The bound region between $P^{A}$ and $P^{B}$ represents the uncertainty, and the mean probability distribution $\overline{P}$ is given by the weighted average, given by
\begin{equation}\nonumber
\begin{split}
\overline{P} &= \frac{P^A/mse^A + P^B/mse^B}{1/mse^A + 1/mse^B} \\
&= \frac{mse^B\,P^A + mse^A\,P^B}{mse^A + mse^B} \\
\end{split}
\end{equation}
\section*{Results \& Discussion}
\begin{figure}[h]
\begin{center}
    \includegraphics[width=0.8\linewidth]{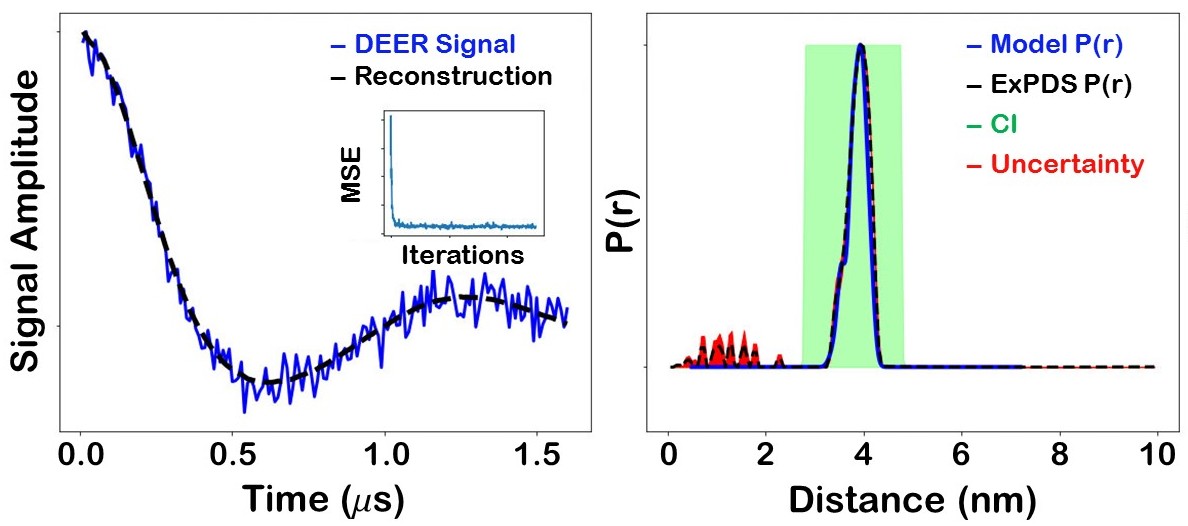}
    \caption[ExPDS P(r) Reconstruction]{(Left panel) Illustration of ExPDS reconstruction (black) of a model DEER time domain signal (blue) with snr of 41. (Right panel) Comparison between the model P(r) (blue) and the mean ExPDS P(r) (black), along with the confidence interval (green box) and the uncertainty (red shaded area).}
  \label{fig:expds_ex1}
\end{center}
\end{figure}

\begin{figure}[h]
\begin{center}
    \includegraphics[width=0.8\linewidth]{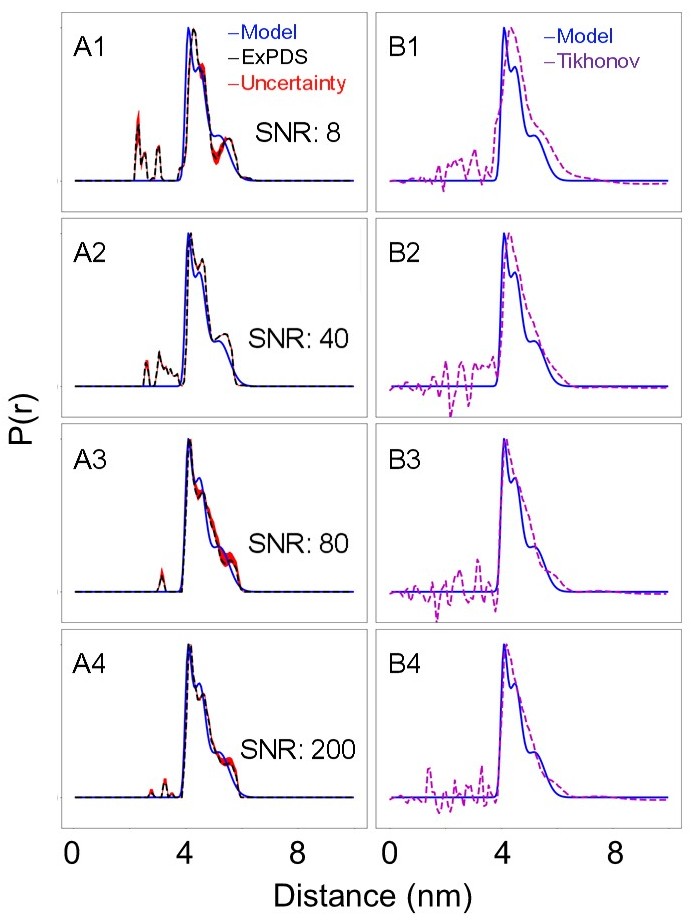}
    \caption[ExPDS TIKR Comparison-I]{Reconstruction of P(r) from model DQC time domain data using (A) ExPDS and (B) Tikhonov regularization with dipolar evolution time and time increment of 4.0 $\mu{s}$ and 32 ns, and increasing snr (1-4). Only the mean ExPDS solution within the confidence interval is shown. The RMSE of fit of the ExPDS and Tikhonov regularized P(r) are (1) 0.1007, 0.1110, (2) 0.0816, 0.0980, (3) 0.0433, 0.0727, and (4) 0.0464, 0.0677.}
  \label{fig:expds_tikr_ex1}
\end{center}
\end{figure}

\begin{figure}[h]
\begin{center}
    \includegraphics[width=0.8\linewidth]{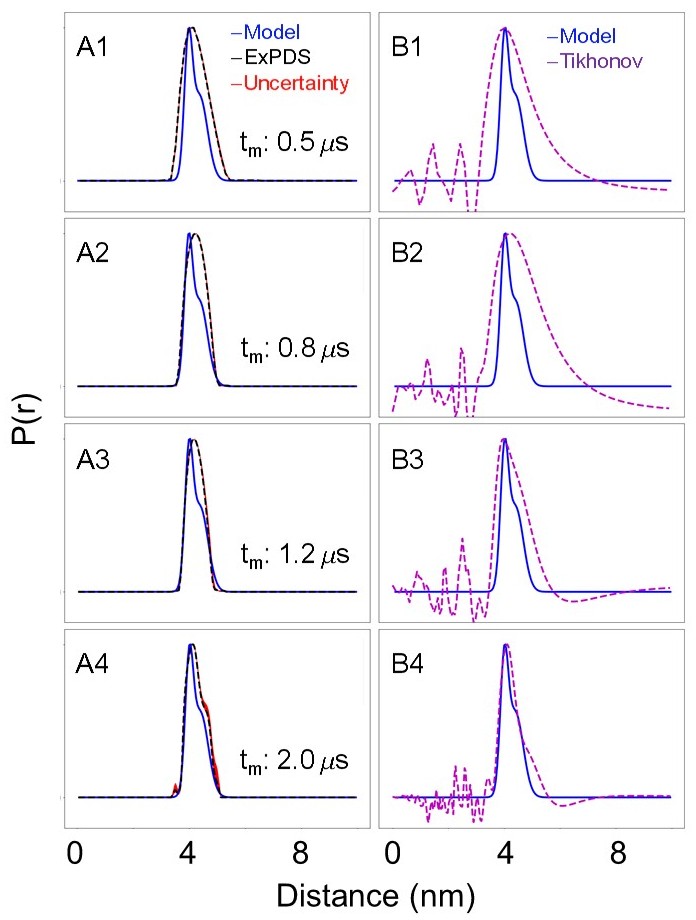}
    \caption[ExPDS TIKR Comparison-II]{Reconstruction of P(r) from model DEER time traces using (A) ExPDS and (B) Tikhonov regularization with dipolar evolution time varying between 0.8 and 2.0 ns, a time increment of 8 ns, and an snr of 100. Only the mean ExPDS solution within the confidence interval is shown. The major peaks in the distribution are located at 4.0 nm, and 4.45 nm. The RMSE of fit of the ExPDS and Tikhonov regularized P(r) are (1) 0.1300, 0.2656, (2) 0.0948, 0.2690, (3) 0.0665, 0.1547, and (4) 0.0513, 0.0750.}
  \label{fig:expds_tikr_ex2}
\end{center}
\end{figure}

We have used synthetic double electron electron resonance (DEER), and double quantum coherence (DQC) data with varying snr, dipolar evolution time, and multi-modal model distance distributions in comparing the efficacy of ExPDS to that of Tikhonov regularization. The distance distribution in \fref{fig:expds_ex1} corresponds to that of between the residue pairs of (5, 132) of T4-Lysozyme \cite{opt_tkn}, a model protein used in PDS studies. Notice that despite a low snr of 41, ExPDS fits the distance distribution with high accuracy (root mean squared error or RMSE of 0.0481) and the artifacts appear outside of the confidence interval, marked by a high level of uncertainty.
\ypar Next we compare the effectiveness of ExPDS and Tikhonov regularization, and the robustness of the solution against varying snr. Both ExPDS and Tikhonov regularization derived distance distributions from a set of synthetic DQC signals are shown in \fref{fig:expds_tikr_ex1} with increasing snr of 8, 40, 80 and 200. The superior fit of ExPDS derived P(r) compared to that of Tikhonov is visible across all the cases, while with increasing snr, the artifacts in the range of 2-4 nm are suppressed significantly in the case of ExPDS. This is a highly desirable feature in order to reduce the time, and cost of signal acquisition.
\xpar We have identified that the dipolar evolution time is another controlling factor in PDS signal analysis, and especially for longer distances (i.e., $r\ge$6 nm), it is often difficult to achieve a sufficiently long $t_m$. In \fref{fig:expds_tikr_ex2}, we compare between ExPDS and Tikhonov regularization in extracting P(r) from a set of synthetic DEER data with varying $t_m$ between 0.8 and 2.0 $\mu{s}$. It should be noted that the corresponding model distance distribution with major peaks at 4.0 nm and 4.45 nm requires a dipolar evolution time of 2.75 $\mu{s}$ or more for optimum resolution \cite{long_range_deer} and hence, $t_m$ value of 0.8 or even 2.0 $\mu{s}$ is significantly lower than the optimal value. Across all the values of $t_m$, ExPDS outperforms Tikhonov regularization in deriving the distance distribution and at 2.0 $\mu{s}$, it provides a near perfect fit to the model distribution (RMSE of 0.0513).

\section*{Conclusion}
We have presented a fully automated, model-free global method for the derivation of distance distributions from PDS time domain signals, which standardises PDS signal analysis. The confidence interval feature of ExPDS identifies the actual range of the distance distribution and the sensitivity analysis provides an uncertainty measure in the solution, which is useful and necessary for refining the derived distance distribution. Along with its robustness against signal noise, the resolution of the ExPDS derived P(r) is significantly better than that of Tikhonov regularization for signals with insufficiently long dipolar evolution time. Therefore, we believe that ExPDS should replace Tikhonov regularization as the global method to be used in combination with specialized local methods, such as SF-SVD in deriving distance distributions from PDS signals and a resemblance between the global, and the local solutions are going to indicate the global minima.
\clearpage
\bibliography{expds}

\end{document}